\documentclass[conference]{IEEEtran}
\usepackage{}
\usepackage{txfonts}
\usepackage{amssymb}
\usepackage{stmaryrd}
\usepackage{caption}
\usepackage{graphicx}
\usepackage{subfigure}
\usepackage{float}

\usepackage{amsmath}
\usepackage{latexsym}
\usepackage{morefloats}
\usepackage{cite}

\usepackage{ifpdf}
\usepackage{algorithmic}
\usepackage{array}
\usepackage{mdwmath}
\usepackage{eqparbox}
\usepackage{mathrsfs}
\usepackage{amsfonts}
\usepackage[]{graphicx,indentfirst}
\usepackage{indentfirst}
\usepackage{slashbox}
\usepackage[justification=centering]{caption}
\usepackage{multirow}
\usepackage{hhline}
\usepackage{bigstrut}
\usepackage{algorithm}    
\usepackage{algorithmic}  
\usepackage{setspace}
\usepackage{yhmath}

\begin{document}

\title{Performance Comparison between Network Coding in Space and Routing in Space}

\author{\IEEEauthorblockN{Yuanqing Ye, Xin Huang, Ting Wen, Jiaqing Huang and Alfred Uwitonze}  
\IEEEauthorblockA{Department of Electronics and Information Engineering,\\
Huazhong University of Science \& Technology (HUST), Wuhan, 430074, P.R. China\\
Email: yuanqingyeah@gmail.com
}
}

\maketitle

\begin{abstract}
Network coding in geometric space, a new research direction also known as $Space~Information~Flow$, is a promising research field which shows the superiority of network coding in space over routing in space. Present literatures proved that given six terminal nodes, network coding in space is strictly superior to routing in space in terms of single-source multicast in regular (5+1) model, in which five terminal nodes forms a regular pentagon centered at a terminal node. In order to compare the performance between network coding in space and routing in space, this paper quantitatively studies two classes of network coding in space and optimal routing in space when any terminal node moves arbitrarily in two-dimensional Euclidean space, and $cost$ $advantage$ is used as the metric. Furthermore, the upper-bound of $cost$ $advantage$ is figured out as well as the region where network coding in space is superior to routing in space. Several properties of $Space$ $Information$ $Flow$ are also presented.

\end{abstract}

\IEEEpeerreviewmaketitle

\section{Introduction}
$Network$~$Information$~$Flow$~(NIF)\cite{NIFNC}, studying $network$ $coding$ $in$ $graphs$, was proposed in 2000. NIF can improve the throughput of a network and reduce the complexity of computing the optimal transmission scheme \cite{LI-LI}. The ratio of maximum throughput of network coding over that of routing is known as $coding$ $advantage$\cite{codingad}.
\par
$Space$~$Information$~$Flow$~(SIF)\cite{LI-LI}, studying $network$ $coding$ $in$ $space$, was proposed by Li $et$ $al$. The $space$ here refers to the geometric space. In this paper, we focus on two-dimensional Euclidean geometric space. In the SIF model, information flows are free to propagate along any trajectories in the space and may be encoded wherever they meet. The purpose is to minimize natural $network$ $volume$, which can support end-to-end unicast and multicast communication demands among terminals in the space, and $network$ $volume$ represents the cost of constructing a network. Taking the unique encoding ability of information flows into account, SIF models the fundamental problem of information network design, which deserves more research attention. The ratio of the minimum routing cost and minimum network coding cost in terms of required throughput is known as $cost$ $advantage$ (CA)\cite{codingad}. CA is used as the metric in the study of SIF. Furthermore,~$cost$~$advantage$~and~$coding$~$advantage$~are~dual.
\par
Yin $et$ $al$.\cite{Yin} studied the properties of SIF, such as Convexity property and Convex Hull property. The literature\cite{Yin} proved that if the number of given terminal nodes is three, CA is always equal to one.  However, the cases where the number of the given terminal nodes is greater than three have not been discussed. Xiahou $et~al$.\cite{Xiahou} proposed a unified geometric framework in space to investigate the Li-Li conjecture on multiple unicast network coding in undirected graphs. Huang $et$ $al$.\cite{2phase} proposed a two-phase heuristic algorithm for approaching the optimal SIF and constructed the Pentagram model. Furthermore, the literature \cite{2phase} proved that the value of CA is 1.0158 in the Pentagram network.
Huang $et~al$.\cite{chongqing} studied the regular (n+1) model in which n terminal nodes formed a regular polygon centered at another terminal node, and proved that only when n=5, network coding in space can be superior to routing in space. The pentagram network is equivalent to the regular (5+1) model. However, the cases where any given terminal node is allowed to move arbitrarily, also called the $irregular$ (5+1) $model$,  have not been discussed. Zhang $et$ $al$.\cite{zhangxiaoxi} discussed the region where CA$\geq$1 in the irregular (5+1) model, when only one terminal node is placed on the vertex of the regular pentagon is allowed to move along the circumcircle. Wen $et$ $al$.\cite{wenting} discussed the region where CA$\geq$1 in the irregular (5+1) model, when only one terminal node that is placed on the vertex of the regular pentagon is allowed to move in space arbitrarily. But the case where the center terminal node is allowed to move arbitrarily as well as the properties associated with this case have not been discussed. This paper studies two classes of irregular (5+1) Model and compares their differences and similarities in order to study the performance of network coding in space and the properties of SIF.
\par
The $contribution$ of this paper is that we quantitatively compare the performance between network coding in space and routing in space through studying two classes of irregular (5+1) Model, and we obtain some properties of SIF, the upper bound of CA and the region where CA$\geq$1.
\par
The organizations of this paper are as follows. Model and definitions are described in Section II. The performance of network coding in space and routing in space are studied in Section III and Section IV, respectively. The numerical analysis and results are presented in Section V. Some properties of network coding in space are discussed in Section VI. Lastly, the conclusions are given in Section VII.
\section{Model and Definitions}
\label{Sec:definition}
\begin{figure}[htp]
\begin{minipage}[t]{0.3\linewidth}
  \centering
  \includegraphics[width=1\textwidth]{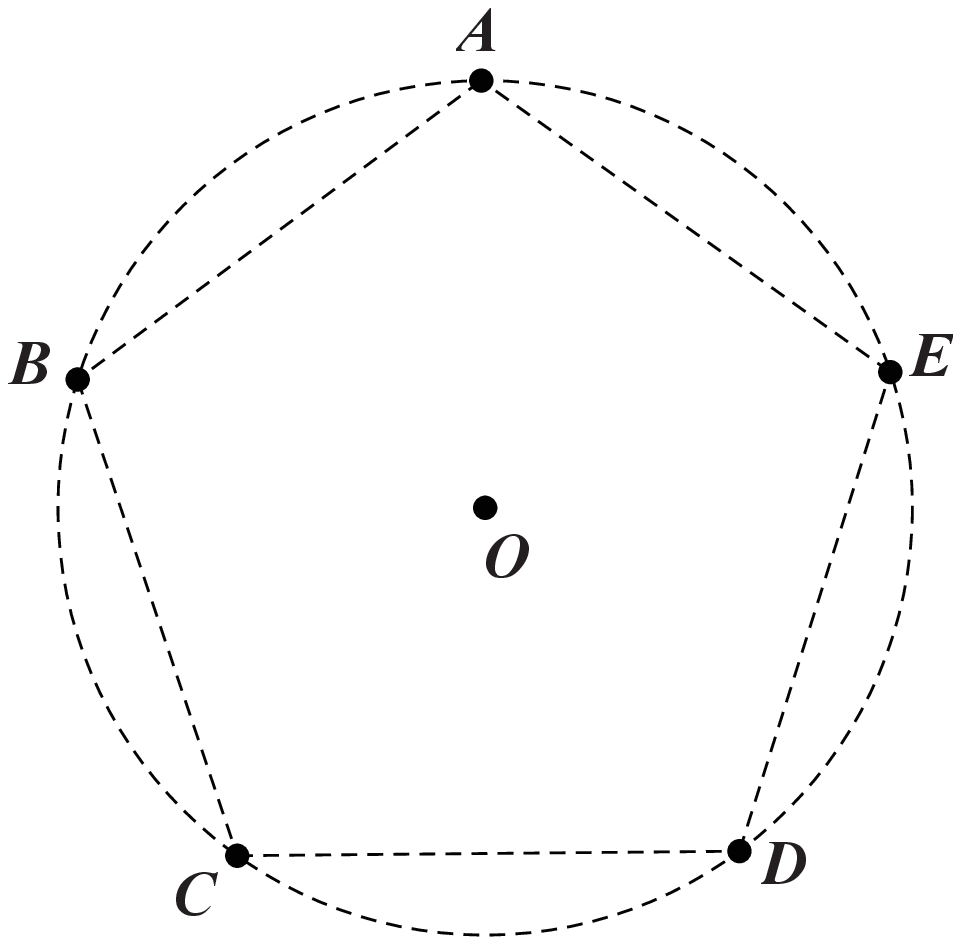}
  \caption{Regular (5+1) model}
  \label{Regular (5+1) model}
\end{minipage}
\begin{minipage}[t]{0.7\linewidth}
\centering
\includegraphics[width=0.84\textwidth]{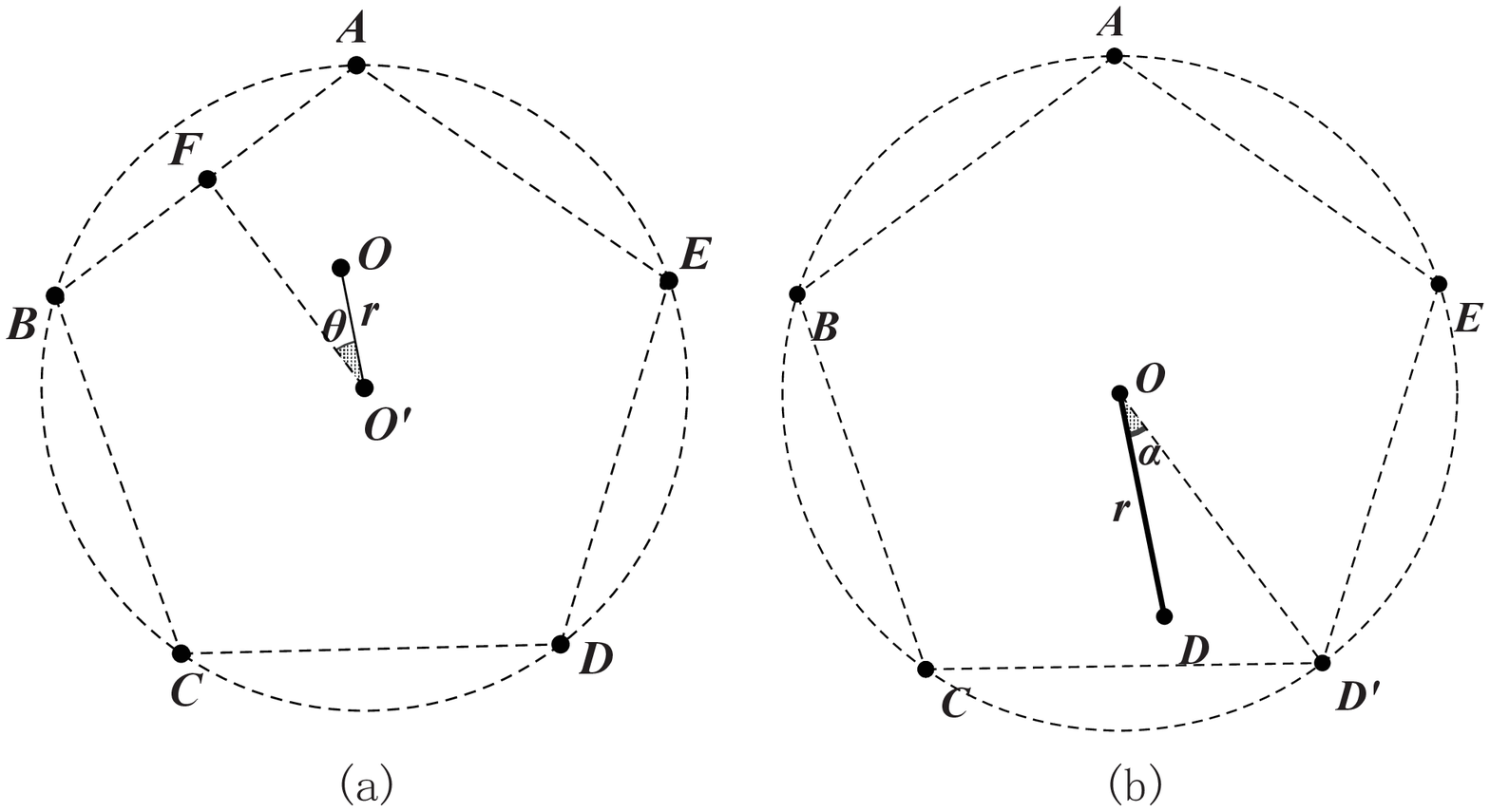}
\caption{Irregular (5+1) model:\\(a) Node Class I; (b) Node Class II}
\label{Irregular (5+1) model}
\end{minipage}
\end{figure}

\newtheorem{definition}{Definition}
 The purpose is to find the min-cost of multicast network coding in space. The cost is defined as $\sum _{e}(\|e\|f_e)$\cite{LI-LI} where $\|e\Arrowvert$ is the length of link $e$\cite{LI-LI} and $f_e$ is the flow rate of link $e$.
\begin{definition}
\textit{(Cost Advantage (CA)\cite{codingad})} CA is defined as the ratio of the minimum routing cost and minimum network coding cost in terms of required throughput. CA is used as the metric to quantitatively compare the performance between network coding in space and routing in space.
\end{definition}
\begin{definition}
\textit{(Regular (5+1) model\cite{2phase})} Given (5+1) terminal nodes in two-dimensional Euclidean space, five terminal nodes $A\sim E$ are the vertices of a regular polygon, whose  circumcenter is the terminal node $O$ (See Fig.\ref{Regular (5+1) model}). The center terminal node $O$ is considered as the source terminal node, and the remaining five terminal nodes are sink terminal nodes.
\end{definition}
\par
\begin{definition}
\textit{(Irregular (5+1) Model)} One of the terminal nodes in the regular (5+1) model deviates from its original position. There are mainly two classifications: Node Class I and Node Class II. The terminal node $O$ denotes the single source terminal node and the terminal nodes $A\sim E$ denote five sink terminal nodes (See Fig.\ref{Irregular (5+1) model}).
\end{definition}
\begin{definition}
\textit{(Node Class I)} The terminal node at the circumcenter is allowed to  move arbitrarily inside the circle. In this Node Class, the terminal node $O$ is the center terminal node that is allowed to move arbitrarily, while node $O'$ is the center of the circumcircle that is fixed. $r$ is depicted as the distance between the center terminal node $O$ and node $O'$, and $\theta$ is denoted as $\angle OO'F$ where $F$ is the midpoint of the line $AB$ (See Fig.\ref{Irregular (5+1) model} (a)).
\end{definition}
\begin{definition}
\textit{(Node Class II)} One of the five terminal nodes on the circumcircle is allowed to  move arbitrarily. In this Node Class, node $D'$ is at the place of one sink terminal node in regular model. $r$ is depicted as the distance between the terminal nodes $O$ and $D$, and $\alpha$ is denoted as $\angle DOD'$ (See Fig.\ref{Irregular (5+1) model} (b)).
\end{definition}
%
\section{Performance of Network Coding in Space}
\subsection{Cost of Network Coding in space for Node Class I}
\begin{figure}[htp]
\centering
\includegraphics[width=0.34\textwidth]{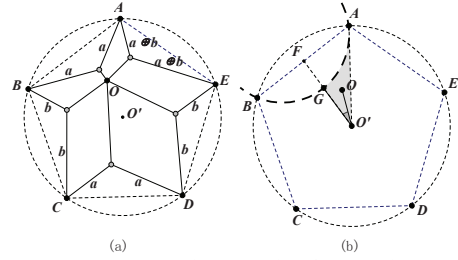}
\caption{Network coding model for Node Class I:\\(a) construction of network coding in space;\\(b) region where network coding in space works}
\label{NC Model for Node Class I}
\end{figure}

The construction of network coding in space is depicted in Fig.\ref{NC Model for Node Class I} (a), and the hollow nodes are relay nodes.
\par
The shaded region where network coding in space works when $0\leq \theta \leq 36\textordmasculine$~and $r\geq0$ is shown in Fig.\ref{NC Model for Node Class I} (b). Only the case that $\theta$ ranges from 0 to 36\textordmasculine ~clockwise is necessarily the one to be discussed about because of the symmetry and $\angle AO'B$= 72\textordmasculine. $\angle BOA$ should be smaller than 120\textordmasculine~according to the Lune property\cite{1968}, otherwise SIF can not help.  As shown in Fig.\ref{NC Model for Node Class I} (b), $\angle AOB\geq$120\textordmasculine~when the terminal node $O$ moves into arc $\wideparen{AGB}$ where $\angle AGB=120\textordmasculine$.
%
\par
Four source flows in space alternately transmit messages $a$ and $b$ while one coding flow transmits the encoded message $a\bigoplus b$, as shown in Fig.\ref{NC Model for Node Class I} (a). Consequently, five sink terminal nodes receive two bits of different messages simultaneously, and $f_e$ =$\frac{1}{2}$ under the assumption that the maximum flow of each sink is unit. Thus the cost of network coding in space can be calculated as follows:

$L_{NC-I}=\frac{1}{2}\times \sum _ {e}(\bigparallel e\bigparallel)=\\
~~~\frac{1}{2}\times\sqrt{r^2-4\sin66\textordmasculine r\cos\theta+4\sin^{2}66\textordmasculine }\\ +\frac{1}{2}\times\sqrt{r^2-4\sin66\textordmasculine r\cos(\theta+72\textordmasculine )+4\sin^{2}66\textordmasculine }\\ +\frac{1}{2}\times\sqrt{r^2-4\sin66\textordmasculine r\cos(\theta+144\textordmasculine )+4\sin^{2}66\textordmasculine }\\ +\frac{1}{2}\times\sqrt{r^2-4\sin66\textordmasculine r\cos(144\textordmasculine-\theta)+4\sin^{2}66\textordmasculine }\\ +\frac{1}{2}\times\sqrt{r^2-4\sin66\textordmasculine r\cos(72\textordmasculine-\theta)+4\sin^{2}66\textordmasculine }$
%
%
\subsection{Cost of Network Coding in space for Node Class II}
\begin{figure}[htp]
\centering
\includegraphics[width=0.34\textwidth]{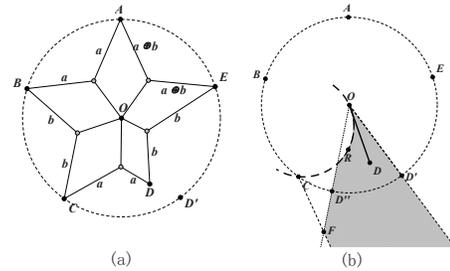}
\caption{Network coding model for Node Class II:\\(a) construction of network coding in space;\\(b) region where network coding in space works}
\label{NC Model for Node Class II}
\end{figure}

The construction of network coding in space is depicted in Fig.\ref{NC Model for Node Class II} (a).
\par
The shaded region where network coding in space works when $0\,^{\circ} \leq \alpha \leq 48\,^{\circ}~$and $r\geq0$ is shown in Fig.\ref{NC Model for Node Class II} (b).
$\angle DOE$, $\angle OCD$ and $\angle CDO$ should be smaller than 120\textordmasculine ~according to the Lune property \cite{1968}, otherwise, SIF can not help. From Fig.\ref{NC Model for Node Class II} (b), $\angle DOE\geq$120\textordmasculine~when the terminal node $D$ moves across the dashed line $OD''$ where $\angle D''OD'=48\textordmasculine$; $\angle OCD\geq120\textordmasculine$~when the terminal node $D$ moves across the dashed line $CF$ where $\angle OCF=120$\textordmasculine; and $\angle CDO\geq120\textordmasculine$ when the terminal node $D$ moves inside arc $\wideparen{ORC}$ where $\angle ORC=120$\textordmasculine.
%
\par
The cost of network coding in space can be calculated as follows\cite{wenting}:\\
$L_{NC-II}=3\cos{24\,^{\circ}}+$\\
$\frac{1}{2}\times \sqrt{1+r ^{2}-2r\cos{(132\,^{\circ}-\alpha )}}+\frac{1}{2}\times \sqrt{1+r ^{2}-\cos{(132\,^{\circ}+\alpha )}}$
\section{Performance of Routing in Space}
\subsection{Cost of Routing in Space for Node Class I}
The cost of routing in space for Node Class I can be obtained by the exact algorithms\cite{steineralgor} of Euclidean Steiner Minimum Tree (ESMT). The main steps are as follows. First, generate all the constructions of  Full Steiner Tree (FST). Second, enumerate all the possible Steiner Trees which are the concatenations of FSTs. Third, calculate the cost of the Steiner Trees and choose the minimum one as ESMT.
\subsubsection{Generation of FSTs }
Generate all the possible FSTs of the six terminal nodes in the irregular (5+1) model.
\subsubsection{Concatenations of FSTs}
\par
\begin{figure}[htp]
\centering
\includegraphics[width=0.40\textwidth]{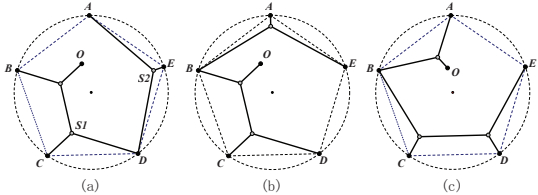}
\caption{Three cases that should be pruned:\\(a) OBCD+DAE; (b) ODCB+BAE; (c) BCDE+ABO}
\label{Three cases pruned}
\end{figure}

We mainly consider the concatenation of one FST with three terminal nodes and another FST with four terminal nodes, and the reasons are similar to\cite{zhangxiaoxi}\cite{wenting}. Moreover, the intersection of FSTs must be the center terminal node $O$ rather than any of the terminal nodes on the circumcircle. All the cases where the intersection of FSTs is the terminal node on the circumcircle can be represented by the three cases shown in Fig.\ref{Three cases pruned}.
According to \cite{1968}, if two FSTs share a node $z$, then the two edges meet at $z$ and make at least 120$\textordmasculine$ with each other. However, when  two FSTs share the terminal node $D$ (See Fig.\ref{Three cases pruned} (a)), $\angle S_1DS_2<\angle CDE=108\textordmasculine<120\textordmasculine$. Hence, Fig.\ref{Three cases pruned} (a) should be pruned. In addition, similar proof can be applied to Fig.\ref{Three cases pruned} (b) and Fig.\ref{Three cases pruned} (c).
%
%
\par
\begin{figure}[H]
\centering
\includegraphics[width=0.40\textwidth]{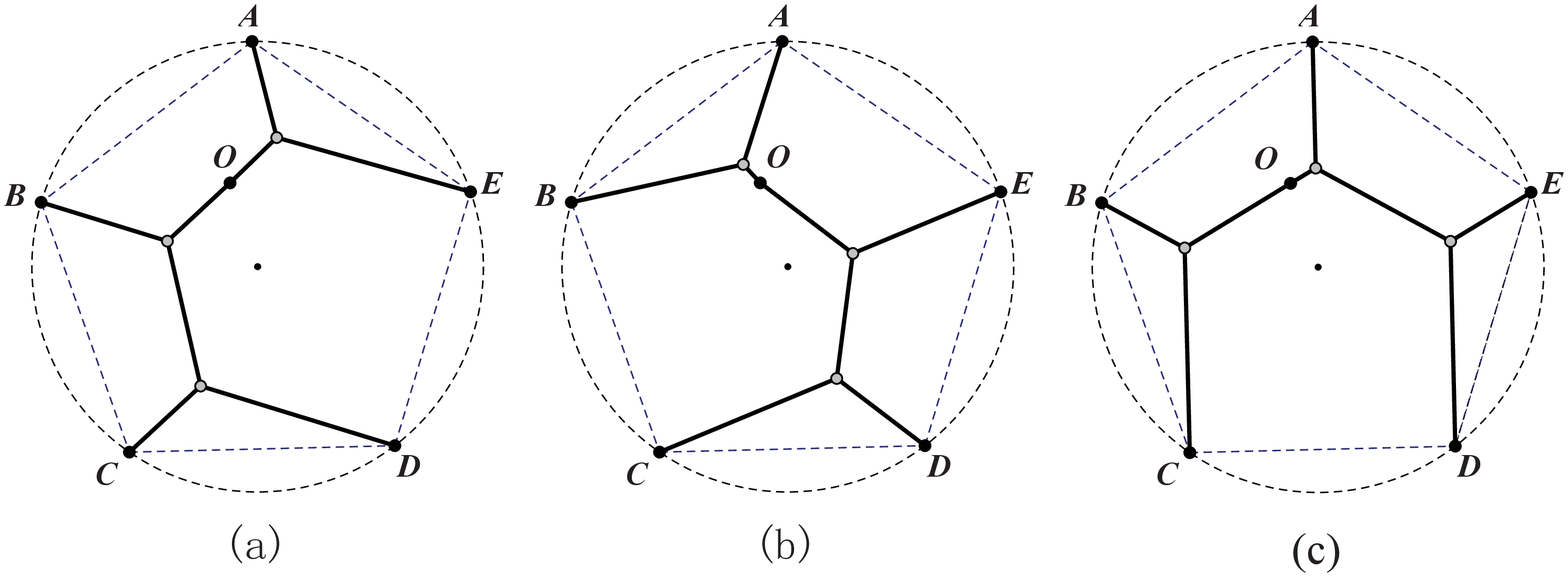}
\caption{First three cases:\\(a) OBCD+OAE; (b) OCDE+OAB; (c) OADE+OBC}
\label{first 3 cases}
\end{figure}
\begin{figure}[H]
\centering
\includegraphics[width=0.40\textwidth]{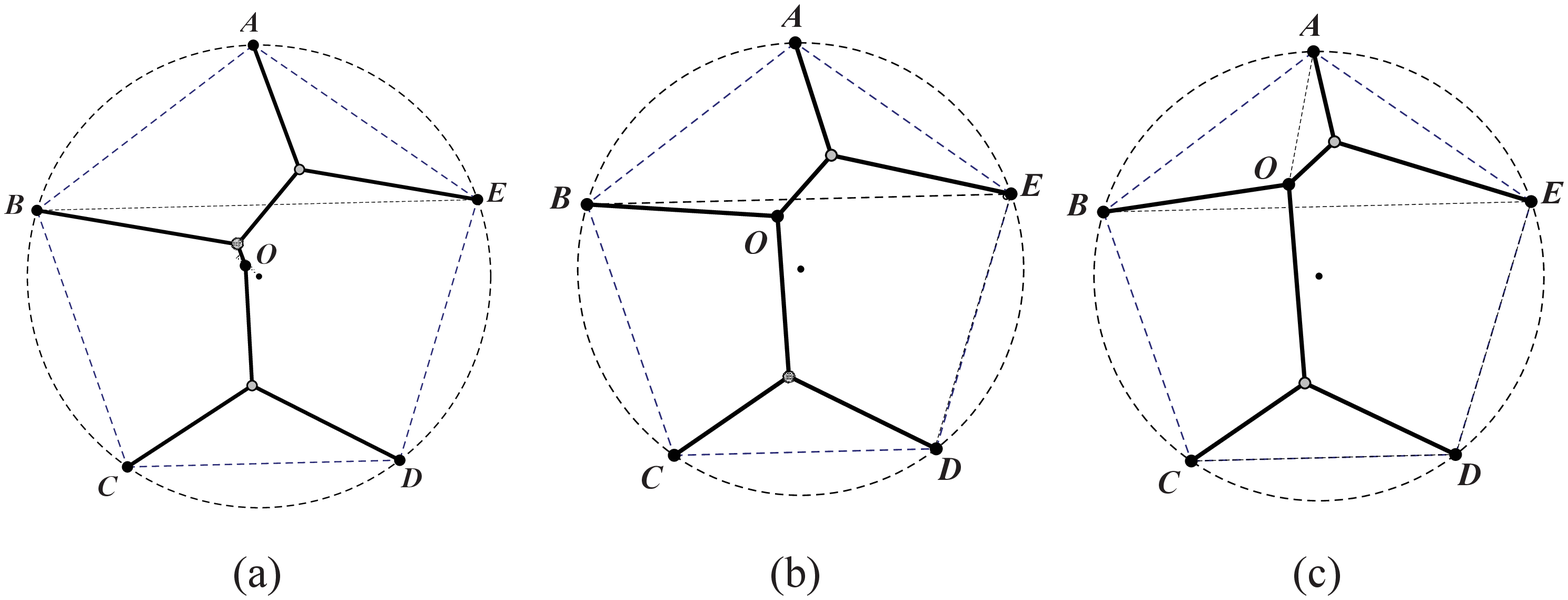}
\caption{The fourth case OABE+OCD has three subcases:\\(a) node O is nondegenerate;\\(b) node O degenerates and is below the line~BE;\\(c) node O degenerates and is above the line BE}
\label{case 4}
\end{figure}
\begin{figure}[H]
\centering
\includegraphics[width=0.40\textwidth]{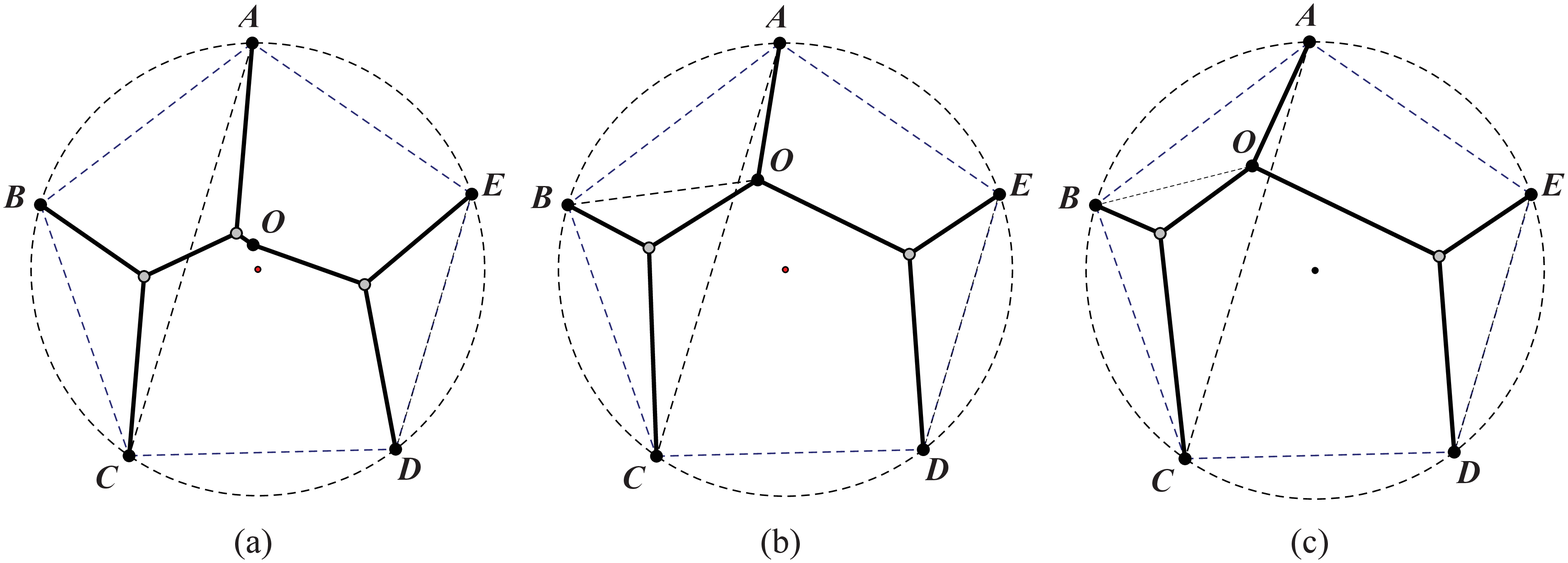}
\caption{The fifth case OABC+ODE has three subcases:\\(a) node O is nondegenerate;\\(b) node O degenerates and is on the right of the line AC;\\(c) node O degenerates and is on the left of the line AC}
\label{case 5}
\end{figure}

The concatenations of FSTs can be divided into five cases, as shown in Fig.\ref{first 3 cases}, Fig.\ref{case 4} and Fig.\ref{case 5}. The subcases shown in Fig.\ref{case 4} (b) and Fig.\ref{case 4} (c) are the degenerations of Fig.\ref{case 4} (a), and the subcases shown in Fig.\ref{case 5} (b) and Fig.\ref{case 5} (c) are the degenerations of Fig.\ref{case 5}~(a).
%
%
%
%
%
\par
\begin{figure}[htp]
\begin{minipage}[t]{0.4\linewidth}
\centering
\includegraphics[width=0.7\textwidth]{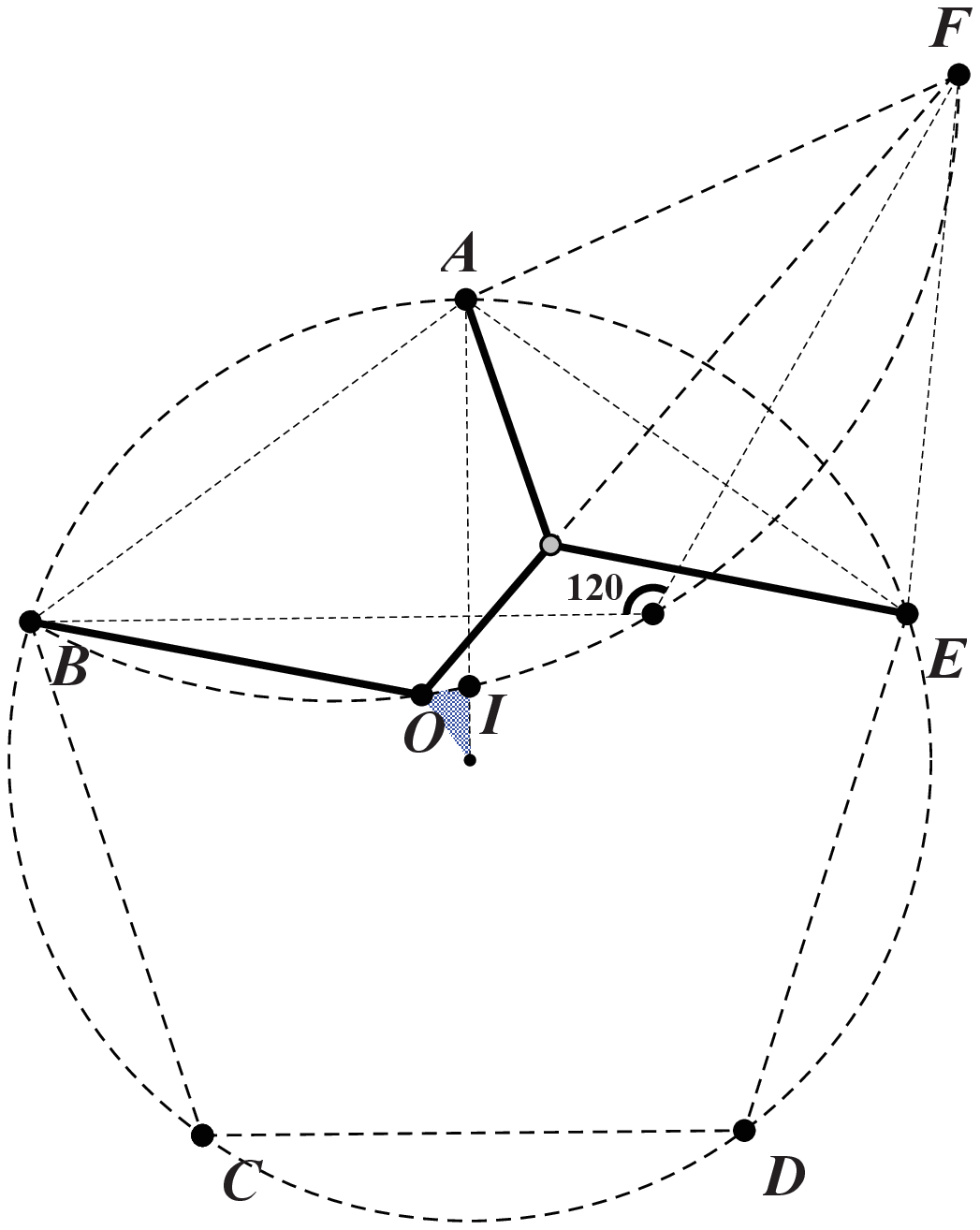}
\caption{Region that the terminal node O does not degenerate}
\label{Case that ESMT does not degenerate}
\end{minipage}
\begin{minipage}[t]{0.5\linewidth}
  \centering
  \includegraphics[width=0.8\textwidth]{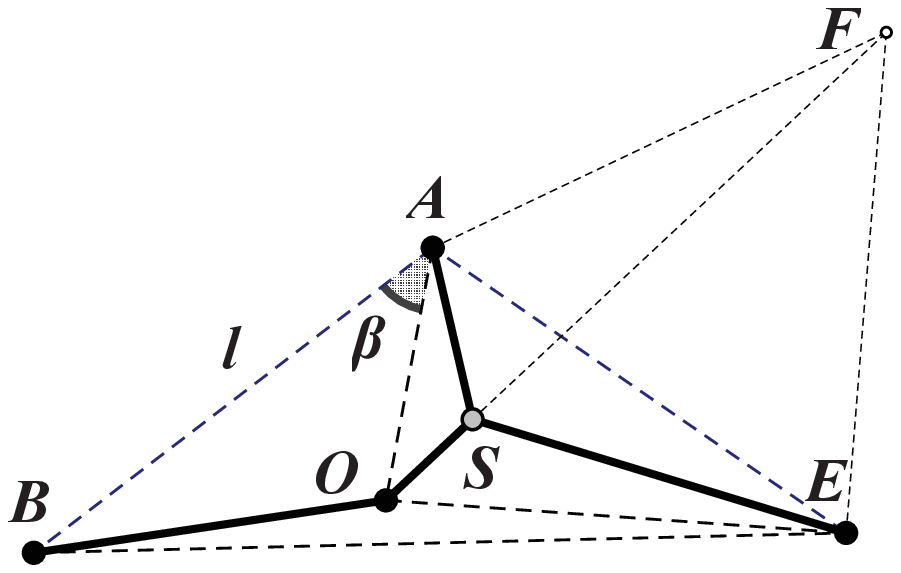}
  \caption{the Steiner node $S$ should lie in $\triangle AOE$}
  \label{S must lie in aoe}
\end{minipage}
\end{figure}
\begin{figure}[htp]
\centering
\includegraphics[width=0.32\textwidth]{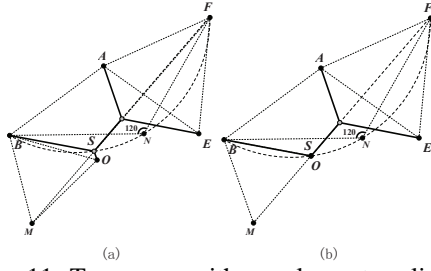}
\caption{Two cases with supplementary lines:\\(a) node $O$ lies outside arc $\wideparen{BNF}$;\\(b) node $O$ is on the arc $\wideparen{BNF}$}
\label{Two cases with supplementary lines}
\end{figure}

In the fourth case, the terminal node $O$ is nondegenerate as shown in Fig.\ref{case 4} (a), when it moves inside the shaded region shown in Fig.\ref{Case that ESMT does not degenerate}. As shown in Fig.\ref{Two cases with supplementary lines}, $\triangle AFE$ and $\triangle BMO$ are two equilateral triangles constructed to find the FST, and the bold-solid lines construct the FST, in which node $S$ is a Steiner node. We construct the arc $\wideparen{BNF}$ that satisfies $\angle BNF$=120\textordmasculine. As every Steiner node of a Steiner Tree has exactly three lines meeting at 120\textordmasculine\cite{1968}, node $S$ is on the arc $\wideparen{BNF}$ when the terminal node $O$ moves outside the arc $\wideparen{BNF}$ as shown in Fig.\ref{Two cases with supplementary lines}~(a). However, when the terminal node $O$ moves on the arc $\wideparen{BNF}$ as shown in Fig.\ref{Two cases with supplementary lines}~(b), $\angle BOF$=120\textordmasculine~and the terminal node $O$ degenerates into a Steiner node. Furthermore, when the terminal node $O$ moves inside the arc $\wideparen{BNF}$, $\angle BOF$$>$120\textordmasculine, and the terminal node $O$ also degenerates.
\par
The terminal node $O$ degenerates as shown in Fig.\ref{case 4}~(b), when it moves below the line BE and is outside the shaded region shown in Fig.\ref{Case that ESMT does not degenerate}. As the terminal node $O$ is below the line BE, terminal nodes $A,~B,~O,~E$ form a convex quadrilateral, and according to\cite{wenting}\cite{1968}, Fig.\ref{case 4} (b) is obtained.
\par
In addition, the terminal node $O$ degenerates as shown in Fig.\ref{case 4}~(c), when it moves above the line BE and is outside the shaded region shown in Fig.\ref{Case that ESMT does not degenerate}. When the terminal node $O$ is above the line BE and forms a concave quadrilateral $ABOE$, the Steiner node should lie in triangle $\triangle AOE$ rather than in triangle $\triangle AOB$. As shown in Fig.\ref{S must lie in aoe}, $\angle AOE<\angle AOB$.
 Let $\angle BAO=\beta$, then $\angle EAO=108\textordmasculine -\beta$. Suppose $\angle AOB\leq120\textordmasculine$, otherwise no Steiner node could possibly exist in $\triangle AOB$\cite{1968}. Let $AO=x$, $AB=AE=l$, then $OE=\sqrt{x^2+l^2-2xl\cos(108\textordmasculine-\beta)}$, $BO=\sqrt{x^2+l^2-2xl\cos \beta}$.\\
(1) If the Steiner node $S$ lies inside $\triangle AOE$, ${L_1}$ denotes the cost of the FST and it is given by:\\ ${L_1}$=$\sqrt{x^2+l^2-2xl\cos(168\textordmasculine-\beta)}$+$\sqrt{x^2+l^2-2xl\cos \beta}$,\\(2) else if the Steiner node $S$ lies inside $\triangle AOB$, ${L_2}$ denotes the cost of the FST and it is given by:\\ ${L_2}$=$\sqrt{x^2+l^2-2xl\cos(\beta+60\textordmasculine)}$+$\sqrt{x^2+l^2-2xl\cos(108\textordmasculine-\beta)}$.

Let $F(\beta)={L_1}-{L_2}$ $(0\leq\beta\leq 54\textordmasculine )$. Calculations show that $F'(\beta)\geq$0, $F(\beta)_{max}$=$F(54\textordmasculine)$ =0. Hence, ${L_1}\leq{L_2}$, and the Steiner node should lie in triangle $\triangle AOE$ and Fig.\ref{case 4} (c) is obtained.
\par
The fifth case has three subcases shown in Fig.\ref{case 5}, similarly to the fourth case.
\subsubsection{Computations of ESMT}
Computations of ESMT are divided into five cases, and ESMT is the one that has the minimum cost.

\begin{figure}[htp]
\centering
\includegraphics[width=0.14\textwidth]{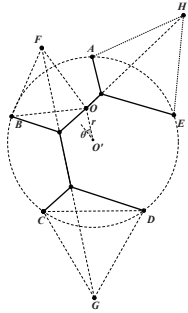}
\caption{Detailed calculation of the first case OBCD+OAE}
\label{Detailed calculation}
\end{figure}

(1)~The first case: $OBCD+OAE$ (See Fig.\ref{Detailed calculation})

%
\par
$\triangle OBF,~\triangle CDG$ and~$\triangle AEH$ are three equilateral triangles, and node $O'$ is the circumcenter, as shown in Fig.\ref{Detailed calculation}.
\par
In $\triangle BO'O$, $\angle BO'O=36\textordmasculine +\theta= \alpha$, $OO'= r$ and $O'A$=1. According to the law of cosines, $BO^2= 1+r^2-2r\cos\alpha$, $\cos(\angle O'BO)= (1-r\cos\alpha )/BO$. In addition, we can obtain $\sin(\angle O'BO)=r\sin\alpha/BO$, ~$\tan(\angle O'BO)= r\sin\alpha/(1-r\cos\alpha)$  by using the law of sines.
\par
In $\triangle BO'G$, $\angle BO'G= 108\textordmasculine$, $O'G= 2\sin66\textordmasculine = \beta$ . Similarly we can obtain $\tan(\angle GBO')=(\beta \sin108\textordmasculine)/(1-\beta \cos108\textordmasculine)$,~$\cos(\angle GBO')= (1-\beta \cos108\textordmasculine)/GB$,~$\sin(\angle GBO')= (\beta \sin108\textordmasculine)/GB$.
\par
In $\triangle FBG$,~$FB^2= 1+r^2-2r\cos\alpha$,~$BG^2 = 1+\beta^2-2\beta\cos108\textordmasculine$,~$\cos(\angle FBG)= \cos(60\textordmasculine+\angle OBO'+\angle O'BG)= 1/(BK\times GB)[0.5-r\sin(\alpha+30\textordmasculine)- \beta\sin138\textordmasculine+ \beta r\sin(138\textordmasculine+\alpha)]$. By using the law of cosines, $\cos\angle FBG= (FB^2+BG^2-FG^2)/(2FB\times BG)$. As a result, $FG^2= 1+r^2+2r\sin(\theta+6\textordmasculine)+ 4r\sin(\theta-6\textordmasculine)\sin66\textordmasculine+ 4(\sin^{2}66\textordmasculine)- 4\sin66\textordmasculine \cos168\textordmasculine$.
\par
In $\triangle O'OH$,~$\angle OO'H= 36\textordmasculine$,~$O'O=r$,~$O'H= \beta= 2\sin66\textordmasculine$. According to the law of cosines,~$OH^2= r^2-4r\sin66\textordmasculine\cos(72\textordmasculine-\theta)+ 4\sin^{2}66\textordmasculine$.
\par
Hence, ${L_{I-1}}=FG+OH$.\\
${L_{I-1}}=[r^2-4\sin66\textordmasculine rcos(72\textordmasculine -\theta)+4\sin^{2}66\textordmasculine ]^{\frac{1}{2}}+$\\
$[1+r^2+2r\sin(\theta+6\textordmasculine) +4\sin66\textordmasculine r\sin(\theta-6\textordmasculine )+4\sin^{2}66\textordmasculine -4\sin66\textordmasculine \cos168\textordmasculine ]^{\frac{1}{2}} $

(2)~The second case: $OCDE+OAB$ (See Fig.\ref{first 3 cases} (b))\\
%
${L_{I-2}}= [r^2-4\sin66\textordmasculine r\cos\theta+ 4\sin^{2}66\textordmasculine]^{\frac{1}{2}}+ [1+r^2-2r\cos(168\textordmasculine -\theta)+ 4\sin66\textordmasculine r\sin(66\textordmasculine -\theta)+ 4\sin^{2}66\textordmasculine -4\sin66\textordmasculine \cos168\textordmasculine ]^{\frac{1}{2}}$

(3)~The third case: $OADE+OBC$ (See Fig.\ref{first 3 cases} (c))\\
$L_{I-3}= [r^2-4\sin66\textordmasculine r\cos(\theta+72\textordmasculine )+ 4\sin^{2}66\textordmasculine ]^{\frac{1}{2}}+ [1+r^2-2r\sin(\theta-6\textordmasculine )- 4\sin66\textordmasculine r\sin(\theta+6\textordmasculine )+ 4\sin^{2}66\textordmasculine -4\sin66\textordmasculine \cos168\textordmasculine ]^{\frac{1}{2}}$

(4)~The fourth case: $OABE+OCD$ (See Fig.\ref{case 4})
\par
When the terminal node $O$ is nondegenerate, then $\angle BOF$$<$120\textordmasculine , i.e. $r\cos(36\textordmasculine - \theta)<\cos72\textordmasculine$ and $\cos(\angle BOF)=$\\
$\frac{1+2r^2-2r\cos(36\textordmasculine+ \theta)+ 4\sin^{2}66\textordmasculine- 4r\sin66\textordmasculine\cos(72\textordmasculine- \theta)- 16\sin^{2}36\textordmasculine\sin^{2}84\textordmasculine}{8\times \sqrt{1+r^2-2r\cos(36\textordmasculine+ \theta)}\times \sqrt{r^2- 4r\sin66\textordmasculine\cos(72\textordmasculine- \theta)+ 4\sin^{2}66\textordmasculine}}>-\frac{1}{2}$,\\
$L_{I-4-1}=[1+r^2-2r\cos(36\textordmasculine+\theta)]^{\frac{1}{2}}+ [r^2-4\sin66\textordmasculine r\cos(72\textordmasculine-\theta)+ 4\sin^{2}66\textordmasculine ]^{\frac{1}{2}}+ [r^2-4\sin66\textordmasculine r\cos(\theta+144\textordmasculine)+ 4\sin^{2}66\textordmasculine]^{\frac{1}{2}}$
\par
When the terminal node $O$ degenerates and is below the line $BE$, then $\angle BOF\geq$120\textordmasculine, i.e.
$r\cos(36\textordmasculine -\theta)<\cos72\textordmasculine$ and $\cos(\angle BOF)=$\\
$\frac{1+2r^2-2r\cos(36\textordmasculine+ \theta)+ 4\sin^{2}66\textordmasculine-4r\sin66\textordmasculine\cos(72\textordmasculine-\theta)- 16\sin^{2}36\textordmasculine\sin^{2}84\textordmasculine}{8\times \sqrt{1+r^2-2r\cos(36\textordmasculine+ \theta)} \times \sqrt{r^2-4r\sin66\textordmasculine\cos(72\textordmasculine- \theta)+ 4\sin^{2}66\textordmasculine}}\leq-\frac{1}{2}$,\\
$L_{I-4-2}=[r^2-4\sin66\textordmasculine r\cos(144\textordmasculine +\theta)+ 4\sin^{2}66\textordmasculine ]^{\frac{1}{2}}+ [1+r^2-2r\sin(66\textordmasculine +\theta)- 4\sin66\textordmasculine r\sin(102\textordmasculine -\theta)+ 4\sin^{2}66\textordmasculine - 4\sin66\textordmasculine \cos168\textordmasculine ]^{\frac{1}{2}}$
 \par
 When the terminal node $O$ degenerates and is above the line $BE$, i.e. $r\cos(36\textordmasculine - \theta)\geq\cos72\textordmasculine$,\\
$L_{I-4-3}=[r^2-4\sin66\textordmasculine r\cos(\theta+ 144\textordmasculine )+ 4\sin^{2}66\textordmasculine ]^{\frac{1}{2}}+ [1+r^2-2r\sin(66\textordmasculine +\theta)- 4\sin66\textordmasculine r\sin(102\textordmasculine -\theta)+ 4\sin66\textordmasculine \cos168\textordmasculine ]^{\frac{1}{2}}$

(5)~The fifth case: $OABC+ODE$ (See Fig.\ref{case 5})
\par
When the terminal node $O$ is nondegenerate: \\
$L_{I-5-1}=[1+r^2-2r\cos(36\textordmasculine -\theta)]^{\frac{1}{2}}+ [r^2-4\sin66\textordmasculine r\cos(\theta+72\textordmasculine )+ 4\sin^{2}66\textordmasculine ]^{\frac{1}{2}}+ [r^2-4\sin66\textordmasculine r\cos(144\textordmasculine -\theta)+ 4\sin^{2}66\textordmasculine ]^{\frac{1}{2}}$
\par
When the terminal node $O$ degenerates and is on the right of the line $AC$:\\
$L_{I-5-2}=[1+r^2-2r\cos(36\textordmasculine -\theta)]^{\frac{1}{2}}+ [r^2-4\sin66\textordmasculine r\cos(\theta+72\textordmasculine )+ 4\sin^{2}66\textordmasculine ]^{\frac{1}{2}}+ [r^2-4\sin66\textordmasculine r\cos(144\textordmasculine -\theta)+ 4\sin^{2}66\textordmasculine ]^{\frac{1}{2}}$
\par
When the terminal node $O$ degenerates and is on the left of the line $AC$:\\
$L_{I-5-3}=[r^2-4\sin66\textordmasculine r\cos(144\textordmasculine -\theta)+ 4\sin^{2}66\textordmasculine ]^{\frac{1}{2}}+ [1+r^2-2r\sin(66\textordmasculine -\theta)- 4\sin66\textordmasculine r\sin(102\textordmasculine +\theta)+ 4\sin^{2}66\textordmasculine -4\sin66\textordmasculine \cos168\textordmasculine ]^{\frac{1}{2}}$
\subsection{Cost of Routing in Space for Node Class II}
Methods are similar with Section IV-$A$. Details refer to \cite{wenting}.
\section{Numerical Analysis and Results}
\subsection{Node Class I}
\par
\begin{figure}[htp]
\centering
\subfigure[]{
\label{Node Class I}
\includegraphics[width=0.22\textwidth]{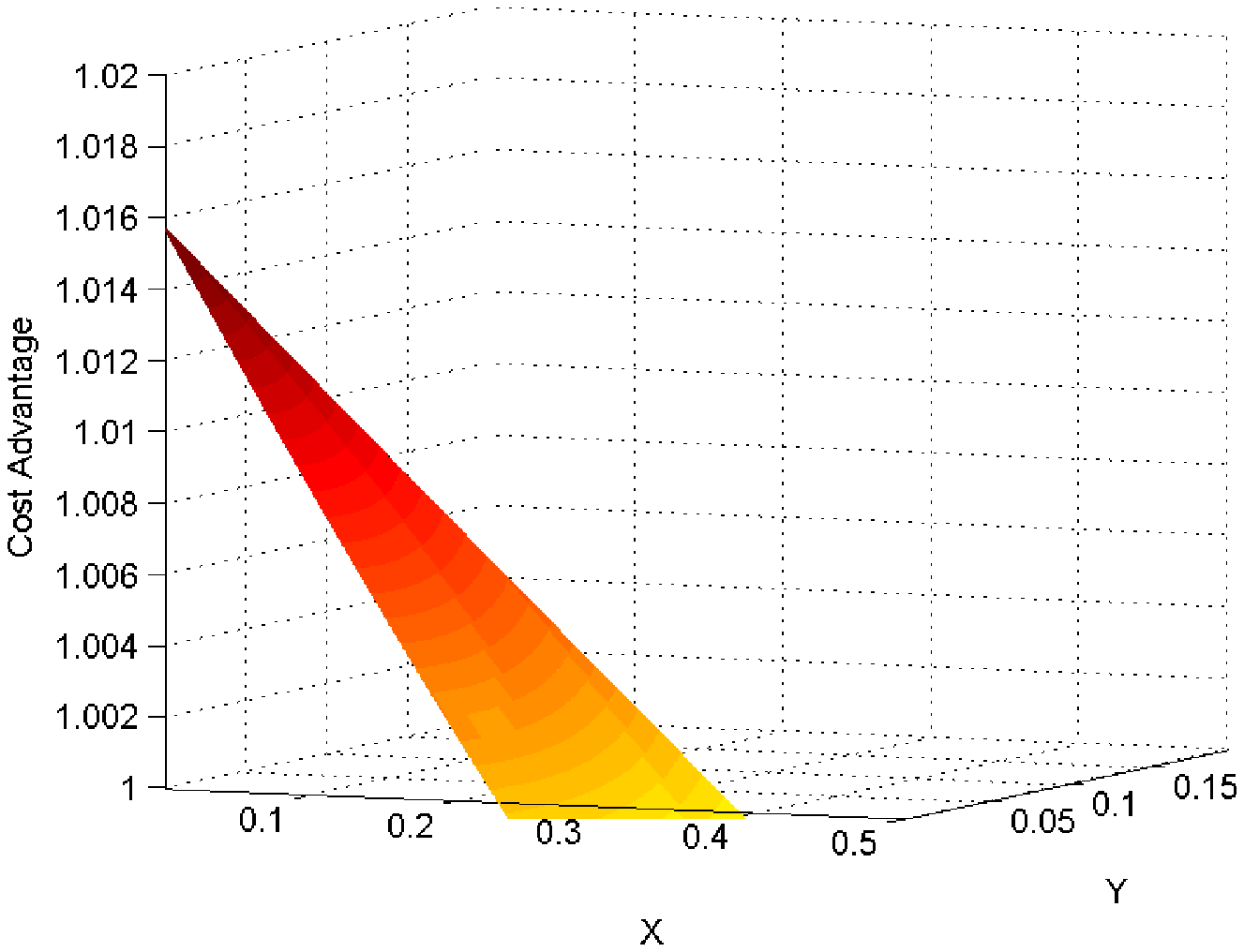}}
\subfigure[]{
\label{Node Class II}
\includegraphics[width=0.24\textwidth]{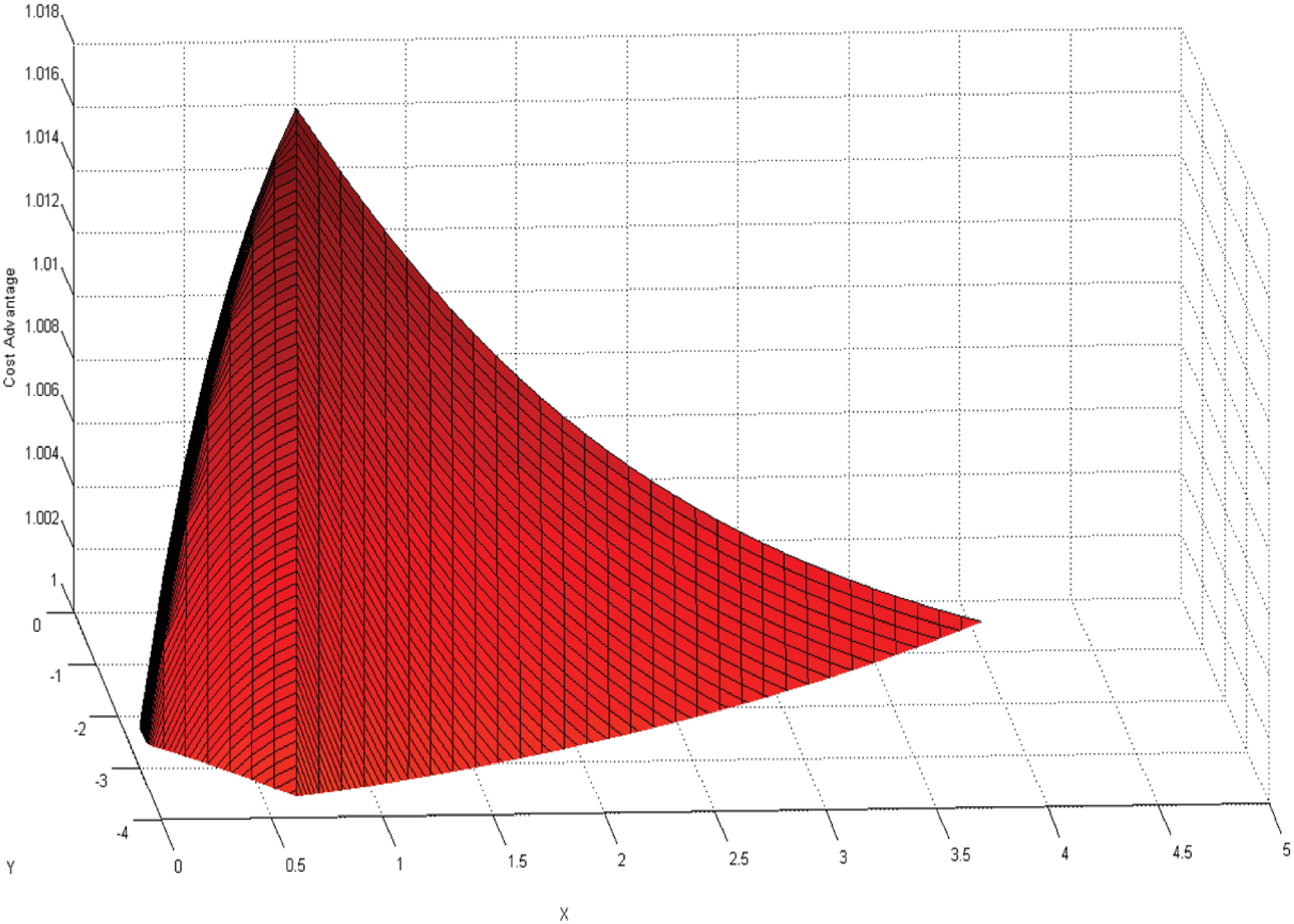}}
\caption{Cost advantage $\geq$1 (3-D):\\(a) Node Class I; (b) Node Class II}
\label{Cost Advantages (3-D)}
\end{figure}

 The functional relation of CA and ($x,y$) in three-dimensional is shown in Fig.\ref{Cost Advantages (3-D)} (a), where only CA$\geq$1 is figured out and cartesian coordinates ($x,y$) are obtained from polar coordinates ($r,~\theta$). Furthermore, Fig.\ref{Cost Advantages (3-D)} (a) shows that CA achieves its maximum value of 1.0158 when $r=0$.
\par
\begin{figure}[htp]
\centering
\includegraphics[scale=.45]{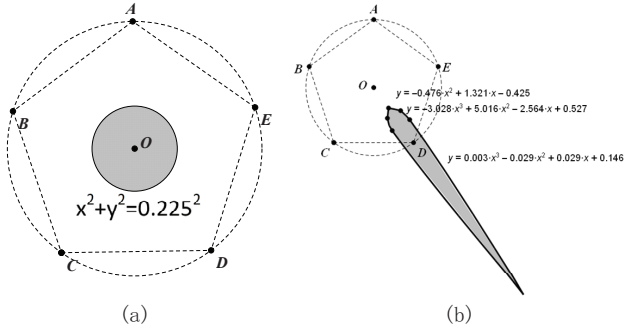}
\caption{Region where cost advantage $\geq$1 (2-D):\\(a) Node Class I; (b) Node Class II}
\label{2D}
\end{figure}

 The two-dimensional region where $CA\geq$1 is shown in Fig.\ref{2D} (a), and it is obtained by projecting Fig.\ref{Cost Advantages (3-D)} (a) to $XY$ plane. When projected to $XY$ plane, Fig.\ref{Cost Advantages (3-D)} (a) turns out to be a sector whose angle  is 36\textordmasculine ~and its radius is between 0.20 and 0.24. Taking the symmetry of the Node Class I model into consideration, the final projection in two-dimensional is shown in Fig.\ref{2D} (a). The performance of network coding in space is superior to routing in space in the shaded region, and the maximum value of CA is achieved when the terminal node $O$ is at the center of the circumcircle (i.e. $r=0$).
%
%
 \par
 Furthermore, it can be proved that the region where network coding in space outperforms routing in space is a circle, and the maximum value of CA can be achieved when the terminal node $O$ is at the centre of the circumcircle. First, transform the problem of CA$\geq$1 to the problem of $L_{NC-I} \leq  min\{L_{I-i}\}~(i=1,2,...,5)$, and make sure there is no discontinuity point in the region where CA$\geq$1. Here $L_{I-4}$ and $L_{I-5}$ represents the minimum value of $L_{I-4-j}~(j=1,2,3)$, and $L_{I-5}$ represents the minimum value of  $L_{I-5-k}~(k=1,2,3)$. Second, when $\theta$ is fixed, study the monotonicity of function $f(r,\theta)=L_{NC-I} - min\{L_{I-i}\}$ in order to confirm that the region where CA$\geq$1 is not an annulus or something similar. Third, when $r$ is fixed, study the monotonicity of function $f(r,\theta)=L_{NC-I} - min\{L_{I-i}\}$ in order to confirm that the region where CA$\geq $1 is a circle.
\par
ESMT should be one of the five cases. Furthermore, the problem of obtaining the region where CA$\geq$1 is equivalent to the the problem of obtaining the region where $L_{NC-I} \leq min\{L_{I-i}\}$.
$L_{NC-I}$ and $L_{I-1}$ to $L_{I-5}$ are all continuous functions, because they are the linear combinations of some basic functions. As a result, function $Y_i=L_{I-i}-L_{NC-I}$ (i=1,2,...,5)~ is also a continuous function. In addition, $f(r,\theta)\geq $0 is a three-dimensional curved surface and as a result its projection on the $XY$ Plane is continuous, which means that there exists no discontinuity point in the region where CA$\geq$1.
 \par
From equations $L_{I-1}~to~ L_{I-5}$, we can calculate it as follows:\\
\par
$\frac{d(L_{NC-I})}{dr}=\frac{0.25}{\sqrt{r^2-4r\sin66\textordmasculine \cos\theta+4\sin^{2}66\textordmasculine }}+\frac{0.25}{\sqrt{r^2-4r\sin66\textordmasculine \cos(\theta+72\textordmasculine )+4\sin^{2}66\textordmasculine }}+\frac{0.25}{\sqrt{r^2-4r\sin66\textordmasculine \cos(\theta+144\textordmasculine +4\sin^{2}66\textordmasculine )}}+\frac{0.25}{\sqrt{r^2-4r\sin66\textordmasculine \cos(144\textordmasculine -\theta)+4\sin^{2}66\textordmasculine }}+\frac{0.25}{r^2-4r\sin66\textordmasculine \cos(72\textordmasculine -\theta)+4\sin^{2}66\textordmasculine }$ ,\\
\par
$\frac{d(L_{I-1})}{dr}=\frac{r+\sin(\theta+6\textordmasculine )+4\sin66\textordmasculine \sin(\theta-6\textordmasculine )}{\sqrt{1+r^2+2r\sin(\theta+6\textordmasculine )+4\sin66\textordmasculine r\sin(\theta-6\textordmasculine )+4\sin^{2}66\textordmasculine -4\sin66\textordmasculine \cos168\textordmasculine }}\\ +\frac{r-2\sin66\textordmasculine \cos(72\textordmasculine -\theta)}{\sqrt{r^2-4\sin66\textordmasculine r\cos(72\textordmasculine -\theta)+4\sin^{2}66\textordmasculine }} $\\
\par
Let $y_1=\frac{d(L_{NC-I})}{dr}-\frac{d(L_{I-1})}{dr}$, when $\theta=\theta_0 ~(\theta_0\in [0,36\textordmasculine ])$, $r\in$[0,0.24]. $r$ is restricted to [0,0.24] according to the projection of Fig.\ref{Cost Advantages (3-D)} (a). By matlab, we find that $\forall r\in[0,0.24]$, $y_1\geq$0. Similar results can be obtained when we calculate the other four functions $y_2=\frac{d(L_{NC-I})}{dr}-\frac{d(L_{I-2})}{dr}$ to $y_5=\frac{d(L_{NC-I})}{dr}-\frac{d(L_{I-5})}{dr}$,  which means that $Y_i=L_{NC-I}-L_{I-i}$ (i=1,2,...,5) are all monotonous increasing. As mentioned above, the function $f(r,\theta)=L_{NC-I} - min\{L_{I-i}\}$ is continuous, thus the function $f(r,\theta_0)=L_{NC-I} - min\{L_{I-i}\}$ is also monotonous increasing when $\theta$ is fixed.
The significance of this result is that if $\exists ~r_0$, when $r=r_0$, $f(r_0,\theta_0)=L_{NC-I} - min\{L_{I-i}\}$=0, then $r_0$ is the only parameter that can satisfy the equation $L_{NC-I} = min\{L_{I-i}\}$. In other words, the region where CA$\geq$1 is not an annulus. In addition, we find that when $r$=0, $CA$=1.0158.
\par
Let $f(r_0,\theta)=L_{NC-I}-min\{L_{I-i}\}$~(i=1,2,...,5), when $r=r_0~(r_0\in[0.20,0.24])$, $\theta\in[0,36\textordmasculine ]$. Calculations show that when $r~(0.20\leq r\leq0.24)$ is fixed, $\frac{df(r_0,\theta)}{d\theta}\geq$0, which means that if $\exists ~r_1\in[0.20,0.24],~f(r_1,0)=0$ and $f(r_1,36\textordmasculine )$=0, then $f(r_1,\theta)$=0 satisfies all of the possibilities when $\theta$ ranges from 0 to 36\textordmasculine, which means the region that satisfies $CA\geq$1 is a circle. Furthermore, $r_1$ indeed exists and $r_1$=0.225 by matlab.
%
\subsection{Node Class II}
The functional relation of CA and ($x,y$) for Node Class II in three-dimensional is shown in Fig.\ref{Cost Advantages (3-D)} (b), where only CA$\geq$1 is depicted, and cartesian coordinates ($x,y$) are obtained from polar coordinates ($r,~\alpha$). Furthermore, Fig.\ref{Cost Advantages (3-D)} (b) shows that CA achieves its maximum value of 1.0158 when ($r,\alpha$)=(1,0). The coordinates of node $D'$  in the irregular (5+1) model is ($r,\alpha$)=(1,0). From Fig.\ref{Cost Advantages (3-D)} (b), the closer the terminal node $D$ moves to the node $D'$, the greater the value of CA gets. CA achieves its maximum value when the terminal node $D$ coincides with the node $D'$.
\par
The two-dimensional region where CA$\geq$1 for Node Class II is shown in Fig.\ref{2D} (b).
\section{Properties and Discussion}
Different from that of routing in space, some properties of SIF can be described as follows:
 (1) Either the center terminal node or  one terminal node on the circumcircle moves arbitrarily, the maximum value of CA is achieved when the irregular (5+1) model turns back to the regular (5+1) model.~
 (2) The number of  relay nodes can be greater than $n-2$ in SIF while this number can not be greater than $n-2$~in ESMT.~
 (3) A given terminal node can have a degree which can be greater than three while the degree can not be greater than three in ESMT.
\par
\begin{figure}[htp]
\centering
\includegraphics[scale=.24]{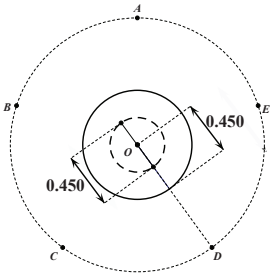}
\caption{Position Independence}
\label{location}
\end{figure}

Furthermore, the center terminal node $O$ (for Node Class I) can only move in the dashed circle shown in Fig.\ref{location}, whose diameter is around 0.450. The minimum distance between one terminal node on the circumcircle and the center terminal node $O$ is around 0.450 (i.e. if the terminal node $D$ for Node Class II moves inside the solid-line circle shown in Fig.\ref{location}, CA will be less than one), which is nearly equal to the diameter mentioned above. Thus, we conjecture Position Independence Property. From Fig.\ref{location}, whatever the position of any terminal node outside the solid circle, or if any terminal node does move anywhere outside the solid circle, the distance between the moving node and the center terminal node $O$ will always be greater than the minimum distance of 0.450 required to achieve CA$\geq$1. The maximum moving distance of the center terminal node is equivalent to the minimum distance between the center terminal node and any other terminal nodes on the circumcircle. Thus, the position of the terminal node does not influence the performance of SIF. In addition, any two terminal nodes can not move too close to each other, otherwise SIF can not help. For example, if the terminal node $O$ moves outside the dashed circle and gets too close to the terminal nodes $A$ and $B$, then it will be too far from the terminal nodes $C$, $D$ and $E$, which will make routing in space superior to network coding in space, meaning that SIF can not help.
\par
 There are following interesting questions to answer: When any two terminal nodes are allowed to move arbitrarily, does CA achieve its maximum value only when the irregular (5+1) model turns back to the regular (5+1) model? When any three terminal nodes are allowed to move arbitrarily, does CA achieve its maximum value only when the irregular (5+1) model turns back to the regular (5+1) model? What if any four terminal nodes are allowed to move arbitrarily? What if any five terminal nodes in this regular (5+1) model are allowed to move arbitrarily, which is equivalent to all the terminal nodes are allowed to move arbitrarily? Furthermore, if given six terminal nodes whose positions are arbitrary, when and how is SIF superior to routing in space? In other words, when given n terminal nodes in space arbitrarily, does SIF help when n=6? Does CA achieve its maximum value only when the six terminal nodes form the regular (5+1) model? It is known that Yin $et~al$\cite{Yin} proposed questions whether SIF can help when n=4 or n=5. Furthermore, Yin has already proved that SIF does not help when n=3.
\section{Conclusions}
This work compares the performance between network coding in space and routing in space based on two classes of irregular (5+1) model, which are Node Class I and Node Class II. Furthermore, CA in both irregular models achieves its maximum value of 1.0158 only when the irregular (5+1) model turns back into regular (5+1) model. The ongoing work is to answer above questions in order to study SIF properties.
\section*{Acknowledgment}
This research was supported by National Natural Science Foundation
of China (No.61271227). The authors thank Zhidong Liu, Wei Xiong and Rui Zhang for their constructive comments.

\begin{thebibliography}{99}
\bibitem{NIFNC}
R. Ahlswede, N. Cai, S.Y.R. Li, and R.W. Yeung. \emph{Network information Flow}.\hskip 1em plus 0.5em minus 0.4em\relax IEEE Trans. on Information Theory, 46(4):1204-1216, 2000.
\bibitem{LI-LI}
Z. Li, B. Li, and L. C. Lau. \emph{On achieving maximum multicast throughput in undirected networks}.\hskip 1em plus 0.5em minus 0.4em\relax IEEE Trans. on Information Theory, 52(6):2467-2485, 2006.

\bibitem{codingad}
S. Maheshwar, Z. Li, and B. Li. \emph{Bounding the coding advantage of combination network coding in undirected networks}.\hskip 1em plus 0.5em minus 0.4em\relax IEEE Trans. on Information Theory, 58(2):570-584, 2012.

\bibitem{Yin}
X. Yin, Y. Wang, X. Wang, X. Xue, Z. Li, \emph{Min-Cost Multicast Networks in Euclidean Space}.\hskip 1em plus 0.5em minus 0.4em\relax IEEE International Symposium on Information Theory (ISIT), 2012.

\bibitem{Xiahou}
T. Xiahou, C. Wu, J. Huang, Z. Li, \emph{A Geometric Framework for Investigating the
Multiple Unicast Network Coding Conjecture}.\hskip 1em plus 0.5em minus 0.4em\relax Netcod, 2012.

\bibitem{2phase}
J. Huang, X. Yin, X. Zhang, X. Du and Z. Li, \emph{On Space Information Flow: Single Multicast}.\hskip 1em plus 0.5em minus 0.4em\relax Netcod, 2013.

\bibitem{chongqing}
J. Huang, F. Yang, K. Jin, Z. Li, \emph{Network Coding in Two-dimension Euclidean Space}.\hskip 1em plus 0.5em minus 0.4em\relax Journal of Chongqing University of Posts and Telecommunications (Natural Science Edition) Oct. 2012.

\bibitem{zhangxiaoxi}
X. Zhang and J. Huang, \emph{Superiority of Network Coding in Space for Irregular Polygons}.\hskip 1em plus 0.5em minus 0.4em\relax IEEE 14th International Conference on Communication Technology (ICCT), 2012.

\bibitem{wenting}
T. Wen, X. Zhang, X. Huang, J. Huang, \emph{Cost Advantage of Network Coding in Space for Irregular 5+1 Model}.\hskip 1em plus 0.5em minus 0.4em\relax IEEE 11th International Conference on Dependable, Autonomic and Secure Computing (ICDASC), 2013.

\bibitem{1968}
E.N. Gilbert and H.O. Pollak. \emph{Steiner Minimal Trees}.\hskip 1em plus 0.5em minus 0.4em\relax SIAM Journal on Applied Mathematics, 16(1): 1-29, 1968


\bibitem{steineralgor}
P. Winter and M. Zachariasen, \emph{Exact Algorithms for Plane Steiner Tree Problems: A Computational Study}. Combinatorial Optimization Volume 6, 2000, pp 81-116.




\end{thebibliography}
\end{document}